\documentstyle[aps,pra,floats]{revtex}

\title{Comments Regarding ``On Neutrino-Mixing-Generated
  Lepton Asymmetry and the Primordial Helium-4 Abundance''}

\author{Xiangdong~Shi, George~M.~Fuller and Kevork~Abazajian}
\address{Department of Physics, University of California,
San Diego, La Jolla, California 92093-0319}

\date{September 10, 1999}

\def\la{\mathrel{\mathpalette\fun <}}

\def\fun#1#2{\lower3.6pt\vbox{\baselineskip0pt\lineskip.9pt
   \ialign{$\mathsurround=0pt#1\hfil##\hfil$\crcr#2\crcr\sim\crcr}}}

\begin{document}
\draft

\twocolumn[\hsize\textwidth\columnwidth\hsize\csname@twocolumnfalse\endcsname

\maketitle

\begin{abstract}
This is a reply to the preprint ``On Neutrino-Mixing-Generated Lepton
Asymmetry and the Primordial Helium-4 Abundance'' by M.~V.~Chizhov and
D.~P.~Kirilova (hep-ph/9908525), which criticised our recent
publication (X.~Shi, G.~M.~Fuller and K.~Abazajian {\sl Phys. Rev.} {\bf D
60} 063002 (1999)). Here we point out factual errors in their description
of what our paper says. We also show that their criticisms of our work
have no merit.
\end{abstract}

\vskip 2pc]
\narrowtext
\noindent{\bf (1)} The main point of the paper by M.~V.~Chizhov and
D.~P.~Kirilova \cite{ck} (hereafter CK) in regards to Shi, Fuller and
Abazajian \cite{sfa} (hereafter SFA) is that the primordial $^4$He
abundance yield in Big Bang Nucleosynthesis (BBN) can be appreciably
affected by neutrino mixing (sterile neutrino production) even when
the lepton number asymmetry, $L$, is small ($L \ll 0.01$, for
example). Of course, this has been known for some time. The change in
the $^4$He abundance yield in an extreme case, $L\rightarrow 0$, was
discussed and calculated as long ago as the early 1980's
\cite{early,barbdolg,enqvist,xsf}. One of the calculations was in fact
done by an author of SFA \cite{xsf}.

On page 2 of the CK paper, it says, ``Certainly such consideration
(meaning that we only consider the $^4$He abundance change for
$L>0.01$) is valid for the simple case of nucleosynthesis {\sl without
oscillations}!''  This is in fact a very good (although not 100\%
accurate) statement regarding the SFA paper. It is obvious that SFA
was indeed only concerned with active-sterile neutrino mixings when
the relevant mixing angles were sufficiently small that the sterile
neutrino production from active-to-sterile neutrino oscillation (other
than the MSW resonant active-to-sterile neutrino conversion whose
amplitude is much less sensitive to mixing angles) is negligible. This
was done for a reason: cases where oscillation effects are large have
been considered before, {\it e.g.}, in the papers cited above.  The
particular parameter space we chose to examine in SFA was based on the
calculation that shows that lepton asymmetry can be generated by
mixings as small as $\sin^2 2\theta\sim 10^{-10}$
\cite{fv,Shi2}. However, oscillation effects (other than MSW resonant
conversion) won't be important until $\sin^2 2\theta \gg 10^{-4}$ for
$\delta m^2 \la 1 \,\rm eV^2$ (see figures of Shi '96\cite{Shi2}). In
this parameter space chosen by SFA, neutrinos or anti-neutrinos can be
converted (via matter-enhanced MSW) to sterile neutrinos, thus
creating a neutrino asymmetry, but the overall neutrino energy density
may not be changed significantly. In such a situation, production of
asymmetries $L\ll 0.01$ indeed {\bf do not} have an appreciable impact
on the primordial $^4$He abundance yield.

This mixing parameter space chosen by us has no overlap with the
parameter spaces considered by CK ($\sin^2 2\theta > 0.01$ and $\delta
m^2<10^{-7}$, from their figure 2, where oscillation effects are
important). It is therefore rather ironic that based on an irrelevant
comparison of two non-overlapping parameter spaces the authors of CK can
claim ``The obtained constraints on $\delta m^2$ are by several orders of
magnitude more severe than the constraints obtained in SFA.''
\bigskip

\noindent {\bf (2)} In the footnote of page 2, CK stated that ``we are
really sorry that...''  Here we are happy to report that the authors of CK
don't have to be sorry because nowhere in the SFA paper did we claim to
be the first to discover this account (of the effects of neutrino spectral
distortion and evolution). The effects of neutrino spectral distortion and
evolution on $^4$He synthesis have been known since the early studies of
BBN, even in the original complete paper on the subject, Wagoner, Fowler
and Hoyle 1967\cite{WFH}. In SFA we merely apply this account to
particular cases of neutrino mixing.  We do not know from which page and
which paragraph in SFA we can be implicated in a claim of discovery.
\bigskip

\noindent {\bf (3)} In regards to the first paragraph of page 8,
the authors of CK are welcome to read more carefully Shi (1996)
\cite{Shi2}, where there is a lengthy discussion on whether the
$L$-generation process meets the classic criterion of chaos (see also
a recent work of Enqvist {\it et al.} \cite{echaos}).  They are also
welcome to produce any evidence showing what will be the sign of a net
lepton number asymmetry $L$ resulting from resonant neutrino
transformation. And yes, even though the chaotic feature of the
$L$-generation process is not well understood, we will ``continue
exploiting it fabricating models and constraints.'' Doesn't any
scientific model involve some assumptions that are not well understood?
We do not believe that our scientific integrity is in any way compromised
when we discuss these models and constraints, because we have always
discussed, and will always continue to do so, the underlying assumptions
of these models and constraints.

Finally, we should point out that the entire problem of
neutrino flavor-transformation in the early universe is a difficult
one. Not the least of the difficulties is solving the Boltzmann
equation plus the MSW equations for multiple particle species with a
spread of energies and occupation numbers. Furthermore, the equations have
non-linear feedback terms that may generate chaos in solutions. Many
groups have attacked these issues. They have obtained many interesting and
important results. But in our opinion, a satisfactory, general solution
has yet to be found. In this sense our understanding of the problem so
far is indeed \lq\lq shallow\rq\rq\ and \lq\lq simplistic.\rq\rq\ We have
no doubt that any future breakthroughs in this problem will offer deeper
and more sophisticated understandings of neutrino physics and cosmology.

\end{document}